# Evaluation of a Novel Quantitative Multiparametric MR Sequence for Radiation Therapy Treatment Response Assessment

**Article Type:** Research Article

Yuhao Yan, M.Sc.,[*, †] R. Adam Bayliss, Ph.D.,[*] Adam R. Burr, M.D., Ph.D.,[*] Andrew M. Baschnagel, M.D.,[*] Brett A. Morris, M.D., Ph.D.,[*] Florian Wiesinger, Ph.D.,[‡] Jose de Arcos Rodriguez, Ph.D.,[§] Carri K. Glide-Hurst, Ph.D.[*, †]

[*]*Department of Human Oncology, University of Wisconsin-Madison, Madison, WI, USA*
[†]*Department of Medical Physics, University of Wisconsin-Madison, Madison, WI, USA*
[‡]*GE HealthCare, Munich, Germany*
[§]*GE HealthCare, Little Chalfont, United Kingdom*

**CORRESPONDENCE**
Carri K. Glide-Hurst, Ph.D., Department of Human Oncology, University of Wisconsin-Madison, 600 Highland Avenue, Madison, WI 53792.
Email: glidehurst@humonc.wisc.edu

**AUTHOR CONTRIBUTIONS**

Yuhao Yan, M.S.: phantom data acquisition, data processing and analysis, manuscript preparation. R. Adam Bayliss, Ph.D.: PI of the HN study, study conception, protocol design, contour delineation, training Yuhao Yan on HN image registrations, manuscript review. Adam R. Burr, M.D., Ph.D.: PI of the HN study, study conception, protocol design, contour delineation, manuscript review. Andrew M. Baschnagel, M.D.: PI of the brain study, study conception, protocol design, manuscript review. Brett A. Morris, M.D., Ph.D.: contour delineation, manuscript review. Florian Wiesinger, Ph.D.: development of DL-MUPA, manuscript review. Jose de Arcos Rodrigues, Ph.D.: development of DL-MUPA, manuscript review.  Carri K. Glide-Hurst, Ph.D.: Co-I of the IRB studies, study conception, manuscript preparation, manuscript review, overall supervision. All authors approved the final version of the manuscript.

**ACKNOWLEDGMENTS**

The authors would like to thank Ana Beatriz Solana, M.S., for her contribution to developing DL-MUPA, Lisa Sanchez for recruiting research subjects and coordinating the clinical studies, MR technicians at UW hospital for acquiring the research scans, and Orhan Unal, Ph.D., for helping with data management. Work reported in this publication was supported by GE Healthcare (PIs: Adam Bayliss, Adam Burr, Andrew Baschnagel. Co-I: Carri Glide-Hurst). Work reported in this publication was supported in part by the National Cancer Institute of the National Institutes of Health under award numbers: R01CA204189 and R01HL153720 (PI: Carri Glide-Hurst). The content is solely the responsibility of the authors and does not necessarily represent the official views of the National Institutes of Health.

## CONFLICT OF INTEREST STATEMENT

Adam Bayliss, Adam Burr, Andrew Baschnagel and Carri Glide-Hurst report research funding from GE Healthcare. Adam Burr reports research funding from Siemens outside of the submitted work. Carri Glide-Hurst reports research collaborations with RaySearch Laboratories and Leo Cancer Care outside of the submitted work (PI: Carri Glide-Hurst). Florian Wiesinger and Jose de Arcos Rodriguez are employees of GE Healthcare. Yuhao Yan and Brett Morris report no conflict of interests.

## DATA AVAILABILITY STATEMENT

Research data are not available at this time.

## Abstract

**Background:** Multi-parametric MRI has shown great promise to derive multiple quantitative imaging biomarkers for treatment response assessment.

**Purpose:** To evaluate a novel Deep-Learning-enhanced MUlti-PArametric MR sequence (DL-MUPA) for treatment response assessment for brain metastases patients treated with stereotactic radiosurgery (SRS) and head-and-neck (HN) cancer patients undergoing conventionally fractionation adaptive radiation therapy.

**Methods:** DL-MUPA derives quantitative T1 and T2 relaxation time maps from a single 4-6-minute scan denoised via DL method using least-squares dictionary fitting. Longitudinal phantom benchmarking was performed on a NIST-ISMRM phantom over one year. In patients, longitudinal DL-MUPA data were acquired on a 1.5T MR-simulator, including pre-treatment (PreTx) and every ~3 months after SRS (PostTx) in brain, and PreTx, mid-treatment and 3 months PostTx in HN. Delta analysis was performed calculating changes of mean T1 and T2 values within gross tumor volumes (GTVs), residual disease (RD, HN), parotids, and submandibular glands (HN) for treatment response assessment. Uninvolved normal tissues (normal appearing white matter in brain, masseter in HN) were evaluated for within-subject repeatability.

**Results:** Phantom benchmarking revealed excellent inter-session repeatability (coefficient of variation <0.9% for T1, <6.6% for T2), suggesting reliability for longitudinal studies with systematic bias adjustment. Uninvolved normal tissue suggested acceptable within-subject repeatability in the brain $|\Delta T1_{mean}|<36$ms (4.9%), $|\Delta T2_{mean}|<2$ms (6.1%) and HN $|\Delta T1_{mean}|<69$ms (7.0%), $|\Delta T2_{mean}|<4$ms (17.8%) with few outliers. In brain, remarkable changes were noted in a resolved metastasis (4-month PostTx $\Delta T1_{mean}$=155ms (13.7%)) and necrotic settings ($\Delta T1_{mean}$=214-502ms (17.6-39.7%), $\Delta T2_{mean}$=7-41ms (8.7-41.4%), 6-month to 3-month PostTx). In HN, two base of tongue tumors exhibited T2 enhancement (PostTx GTV $\Delta T2_{mean}$>7ms (12.8%), RD $\Delta T2_{mean}$>10ms (18.1%)). A case with nodal disease resolved PostTx (GTV $\Delta T1_{mean}$=-541ms (-39.5%), $\Delta T2_{mean}$=-24ms (-32.7%), RD $\Delta T1_{mean}$=-400ms (-29.2%), $\Delta T2_{mean}$=-25ms (-35.3%)). Parotids (PostTx $\Delta T1_{mean}$>82ms (12.4%), $\Delta T2_{mean}$>6ms (13.4%)) and submandibular glands (PostTx $\Delta T1_{mean}$>135ms (14.6%), $\Delta T2_{mean}$>17ms (34.5%)) adjacent to gross disease exhibited enhancement while distant organs remained stable.

**Conclusions:** Preliminary results suggest promise of DL-MUPA for treatment response assessment and highlight potential endpoints for functional sparing.

## Introduction

MRI plays a vital role in radiation therapy (RT) for its excellent soft tissue contrast and increasing availability of MR-simulators (MR-SIMs) sited in Radiation Oncology. One of its key applications is in assessing treatment response[1] through quantitative MR imaging (QMRI) biomarkers. These biomarkers quantify the changes in both physical and functional characteristics, supporting more informed clinical decision-making.[2,3]

In the brain, post-Gadolinium T1 weighted and FLAIR T2 weighted MRIs are often used in the management of metastases. However, conventional qualitative MRIs offer limited ability to distinguish between radiation necrosis, tumor progression, and recurrence, presenting a persistent clinical challenge.[4] Diffusion weighted imaging and perfusion weighted imaging have been studied for assessing radiation treatment response of brain metastases. These techniques enable the measurement of QMRI parameters such as the apparent diffusion coefficient (ADC) and relative cerebral blood volume.[5] Ding *et al*. prospectively assessed post-treatment T1 maps of 56 brain metastases patients treated by Gamma Knife radiosurgery and found the difference between quantitative T1 relaxation time 5-minute and 60-minute post contrast administration sensitive to differentiate tumor recurrence from radionecrosis.[6]

In Head and Neck (HN), to accommodate drastic tumor and anatomical changes that occur during the treatment course, recent advancements of adaptive radiation therapy (ART) strategies have demonstrated dosimetric benefits of organs at risk (OAR) sparing and potential clinical benefits in terms of survival and toxicity.[7] ART allows mid-RT MR acquisitions and subsequent longitudinal QMRI biomarker analysis which can potentially enhance response assessment and guide the adaptation.[8] Mohamed *et al*. prospectively evaluated 81 HN cancer patients treated with conventionally fractionated RT and identified changes of mean ADC values in primary tumor volume greater than 7% at the time of mid-RT (~18 RT fractions) compared to pre-RT as a statistically significant QMRI biomarker associated with better local control and recurrence-free survival.[9] Bonate *et al*. prospectively analyzed 15 HN squamous cell cancer (SCC) patients treated with hypofractionated ART on a 1.5T MR-linac and found significant longitudinal increase in ADC and T2 values of primary tumor over the cohort.[10]

With the promising application of QMRI in RT response analysis, it is appealing to have quantitative multi-parametric MRI (mpMRI) to provide comprehensive functional information, however routine application of mpMRI is currently limited by the need to acquire multiple MRIs with long acquisition time to achieve clinically acceptable resolution,[11] particularly for conventional T1 and T2 quantification methods such as variable flip angle (VFA) and inversion recovery (IR). Recently, Nejad-Davarani *et al*. optimized and implemented a novel mpMRI sequence, STAGE (STrategically Acquired Gradient Echo), on patients with glioblastoma treated on a 0.35T MR-linac with conventional fractionation RT, which acquired quantitative proton density (PD), R2* and T1 maps in 10 minutes. This pilot study observed changes of above QMRI values in primary tumor volume at follow up (2 months) agreeing to findings on diagnostic images.[12] The technique of MR Fingerprinting (MRF) has recently emerged, which simultaneously derives multiple QMRI values in a single scan, matching unique signal evolution of different tissue under the sequence with continuously varying parameters to a pre-calculated dictionary.[13] Given the promising efficacy, efforts have been devoted to clinical translation of mpMRI to facilitate ART and response assessment, focusing on the evaluation of accuracy, repeatability and *in vivo* feasibility.[14,15] In the present work, a vendor-developed deep-learning-enhanced multi-parametric MR sequence, 'DL-MUPA', was implemented that yields 5 qualitative datasets and derives quantitative T1 and T2 relaxation time maps from a single 4-6-minute scan. Compared to commonly acquired quantitative MRI sequences such as DWI, DL-MUPA derives relaxometry parameters, providing complementary physiological information. DL-MUPA also improves clinical efficiency by eliminating the need of multiple acquisitions and shortening the overall scan time. MRF offers several advantages such as more flexible pulse sequence design and simultaneous derivations of comprehensive biomarkers including T1, T2 and diffusion, thus promising for a wide variety of clinical applications.[13,16] However, it faces several challenges such as limited volume coverage and computationally expensive post-processing.[13,16] DL-MUPA on the other hand is already integrated into the vendor platform and reconstruction pipeline leveraging deep learning to improve image quality, allowing straightforward and practical clinical implementation. Accuracy and repeatability were first assessed in a comprehensive phantom study and results were demonstrated in two patient cohorts: brain metastases patients undergoing stereotactic radiosurgery (SRS) and HN cancer patients undergoing conventionally fractionated ART.

## Methods

### Deep-Learning-enhanced MUlti-PArametric MR Sequence – DL-MUPA

DL-MUPA (GE Healthcare, software versions 29.1-30.1, Milwaukee, WI) sequence acquires 1 PD, 3 T1 and 1 T2 weighted qualitative 3D images in a single 4–6-minute scan and derives quantitative T1 and T2 relaxation time maps using least-squares, orthogonal matching pursuit[17] dictionary fitting relative to spoiled gradient-echo (SPGR) simulated signals.[18,19] The sequence starts with a steady-state PD weighted image acquisition using low flip angle (FA, ~1°) Zero TE, followed by a transient-state T1 and T2 magnetization prepared 4-segment Zero TE acquisition (FA=3°) similar to the 3D-QALAS approach, where one acquisition was performed after the T2 preparation, then three acquisitions were performed at different points of the T1 relaxation after the T1 preparation.[20] Images were reconstructed using a DL-based reconstruction pipeline, which uses a deep convolutional neural network (CNN) to reduce noise, ring artifacts and improve sharpness.[21,22] The minimal gradient switching intrinsic to Zero TE has been characterized as "silent" that noise is minimized during acquisition.[23]

### Phantom Benchmarking

#### *Phantom and Data Acquisition*

A NIST-ISMRM quantitative MR phantom (Serial Number 0130, CaliberMRI, Boulder, CO) was scanned to benchmark the DL-MUPA sequence. The phantom contains arrays of 14 sphere vials of $NiCl_2$ and $MnCl_2$ solutions at different concentrations for T1 and T2, respectively, spanning across T1 values of 22-1741 ms and T2 values of 7-493 ms per manufacturer reference, calculated based upon a NIST-provided methodology at 1.5T at 20°C. Following guidance from the manufacturer, two T2 vials (index 1 and 5) were excluded from formal analysis as reported elsewhere.[24,25]

Phantom benchmarking was performed following Quantitative Imaging Biomarker Alliance (QIBA) guidelines.[2,3] The phantom was scanned with DL-MUPA sequence using a 19-channel brain coil (GE GEM head and neck suite) on a GE 1.5T SIGNA Artist MR which is dedicated to radiation oncology use. A single DL-MUPA scan took 4 min 30 sec with 1.6 $mm^3$ acquisition voxel size, 2 averages, 202 $mm^3$ FOV covering the whole phantom. To test the accuracy and repeatability of the sequence, 12 separate phantom imaging sessions were acquired over one year (mean interval ± standard deviation: 34 days ± 35 days, range: [9, 120] days) with 5 consecutive repeat scans at each scanning session. The phantom was stored in the scan room to maintain thermal equilibration and phantom temperature was measured before and after each session to assess the need for temperature corrections.

### *Image Processing and Quantitative Analysis*

An automated MATLAB algorithm was devised to segment the vials on PD weighted images as the contrast and image quality provided definite boundaries of the vials. Each vial was measured by the mean intensity of voxels within 5 mm radius to avoid partial volume effects. Physiologically relevant T1 (246-1741 ms) and T2 (42-493 ms) values were assessed. Accuracy was evaluated by bias comparing mean derived values of all acquisitions (12×5) to manufacturer-provided reference, in both absolute value and percentage relevant to the reference. Repeatability was measured by coefficient of variation ($CV = \frac{Standard\ Deviation}{Mean}$). For each vial, intrasession CV was calculated over the 5 consecutive acquisitions in a single session, thus rendering 12 intrasession CVs, of which only the range was reported. Mean T1 and T2 of each session was calculated averaging the five acquisitions. Intersession CV was calculated over the 12 session means. Linearity between the derived values against the reference were assessed using least squares linear regression via Scipy package in python.

## *In vivo* Feasibility Study

### *Patient Cohort and Data Acquisition*

To study the feasibility of using DL-MUPA derived quantitative T1 and T2 maps to assess treatment response *in vivo*, longitudinal DL-MUPA images were prospectively acquired on IRB-approved protocols to enable experimental longitudinal imaging, one for patients with brain metastases (IRB UW 2021-1013) and one for patients with HN cancer (IRB UW 2022-0321). To demonstrate the feasibility and reliability of using DL-MUPA derived quantitative T1 and T2 maps for *in vivo* treatment response assessment, a subset of the patients was analyzed. Specifically, in the brain metastases cohort, three patients with 8 tumors were evaluated for tumor response and uninvolved normal tissue (normal appearing white matter (NAWM)) repeatability. In the HN cancer cohort, three patients were evaluated for response of gross tumor volumes (GTVs, n=4) and nearby organs at risk (parotids and submandibular glands), with five HN patients evaluated for normal tissue (uninvolved masseter) repeatability.

Patients with brain metastases underwent SRS to 18-21 Gy. The three patients included in this work were all diagnosed with brain metastases from non-small cell lung cancer. Two were female and one was male (age: 71±9 years). Pre-contrast DL-MUPA scans were acquired prior to SRS (PreTx) and at follow-ups every ~3 months after SRS procedure (PostTx). Patients were scanned on the GE 1.5T SIGNA Artist MR with the 19-channel brain coil, identical to phantom benchmarking. DL-MUPA took 4 min 30 sec with 1.5 $mm^3$ acquisition voxel size, 192 $mm^3$ FOV covering the whole brain, 2 averages. Post-contrast T1 weighted images (repetition time (TR)~9 ms, echo time (TE)~3.5 ms, inversion time (TI)=450 ms, FA=13°) and T2 FLAIR (TR=6000 ms, TE~135 ms, TI~1730 ms, FA=90°, echo train length (ETL)=200) were also acquired at each timepoint, which are radiology sequences optimized for stereotactic radiosurgery following our local clinical practices. CT simulation was performed on Siemens SOMATOM Definition Edge (120 kV) the same day as PreTx MR acquisition.

Patients with HN cancer underwent conventional fractionation ART to 70 Gy with plan adaptation performed after the $10^{th}$-$15^{th}$ RT fraction to accommodate anatomical changes. Out of the five patients included in this work, four were diagnosed with SCC and one with adenoid carcinoma. Four were male and one was female (age: 51±7 years). Pre-contrast DL-MUPA scans were acquired at the time of MR-SIM prior to treatment (PreTx), 2-3 weeks after RT start (MidTx), and at 3-month follow-up after treatment end (PostTx). Patients were scanned in treatment position with their immobilization masks on the GE 1.5T SIGNA Artist MR with a 32-channel Head and Neck coil (GE AIR Open RT suite, Fig 1A) or a 30-channel Head and Neck coil (GE GEM RT Open suite, Fig 1B). For HN, DL-MUPA was acquired in 6 min 30 sec with $1.5\times1.5\times2$ $mm^3$ acquisition voxel size, 264 $mm^3$ FOV covering from the apex of lung to the middle level of the brain, 1.5 averages. Pre-contrast 2D T2 frFSE (TR~9000 ms, TE~77 ms, ETL=19, FA=160°) and post-contrast T1 LAVA-Flex (TR~6 ms, TE~3 ms, FA=12°) were also acquired at each timepoint. The former served as a research sequence for auto-contouring tasks while the latter was acquired for radiology review and contouring purposes. PreTx CT simulation was performed on the Siemens SOMATOM Definition Edge (120 kV) 3.4±3.3 days (range: [0, 10] days) from MR-SIM. In addition, Quality of Life (QOL) data was collected pre-treatment, at the end of treatment, and 3-months after treatment using validated a validated QOL instrument (European Organization for Research and Treatment of Cancer Quality of Life (EORTC) Core Questionnaire HN module (QLQ-H&N35)).[26]

### *Image Processing and Delta Analysis*

To demonstrate the feasibility of DL-MUPA for *in vivo* treatment response assessment, three brain metastases patients underwent delta analysis across 8 tumors, and three HN cancer patients underwent delta analysis of four GTVs and nearby organs at risk including parotids and submandibular glands. To identify consistent regions of interest at different timepoints, contouring and image registrations were performed using MIM Maestro v7.2.8 (MIM software, Cleveland, OH). Clinically approved contours were initially delineated on the treatment planning CTs (TPCTs). For the brain, longitudinal DL-MUPA T1 weighted images were rigidly co-registered to TPCT using box-based assisted alignment followed by manual adjustments considering the rigidity of skull and generally consistent anatomy inside the brain.[12] GTVs were then rigidly propagated from TPCT to DL-MUPA images. In HN, longitudinal T2 frFSE images were co-registered to TPCT via multi-modality deformable registration (DIR) followed by manual adjustments locally near the lesion. DIR for HN is used in our clinical practice and others[9,27] due to anatomical variations in patient positioning, tumor changes, and patient weight loss. DL-MUPA T1 weighted images were rigidly co-registered to T2 frFSE images at corresponding timepoints using localized box-based alignment.[9] GTVs were propagated to longitudinal T2 frFSE and contours were finalized by a board-certified HN radiation oncologist. As substantial GTV changes were observed over the treatment course, the MidTx and PostTx Residual Disease (RD) volumes were delineated using information from the T2 frFSE and post-contrast T1 LAVA-Flex following a method defined by Mohamed *et al.*[9] The clinically used parotid and submandibular gland volumes were deformably propagated to the other T2 frFSE timepoints and modified to match the underlying anatomy.[28] All finalized contours were then rigidly propagated from T2 frFSE to DL-MUPA images for quantitative T1 and T2 analysis.

To evaluate the *in vivo* repeatability of DL-MUPA, delta analysis of uninvolved tissue was performed for three brain metastases patients and five HN cancer patients. To define uninvolved tissue as a control for longitudinal datasets, in brain, the normal appearing white matter (NAWM)[12] was derived by subtracting the lesion from white matter automatically segmented on T1 weighted and T2 FLAIR images using SPM12 and LST toolbox.[29,30] Analysis was then conducted based on the initial timepoint at the central plane of the GTVs, in the uninvolved contralateral or ipsilateral hemisphere, then propagated to the secondary timepoint via rigid registration for longitudinal assessment. For HN, a single slice of uninvolved contralateral masseter[31] was manually delineated on PreTx images as a control volume.

To mask out the background, Otsu thresholding[32] was performed on the DL-MUPA-derived PD weighted images. In addition, in HN, the airway was carefully masked out by implementing seed fill algorithm. For quantitative analysis, mean T1 and T2 values within each region of interest (ROI) were calculated at each timepoint. Delta analysis was performed calculating longitudinal changes of mean T1 and T2 values with respect to PreTx for ROIs including GTV, parotids and submandibular glands for HN, and uninvolved normal tissue (contralateral NAWM in Brain and masseter in HN).[12,31,33,34] In addition, for HN, major anatomical changes were addressed by isolating the residual disease and comparing mean T1 and T2 values within MidTx and PostTx RD to PreTx GTV (initial disease) via histogram analysis. [9] Furthermore, for each patient, within-subject coefficient of variation ($wCV = \frac{Standard\ Deviation}{Mean}$) was calculated for each ROI across all analyzed timepoints.

## Results

### Phantom Benchmarking

Table 1 summarizes the quantitative results of phantom benchmarking across 12 sessions. Mean derived T1 and T2 values showed systematic bias of ~25-35% compared to manufacturer reference. By contrast, intersession CVs were minimal within 0.9% for all T1 values and within 2.4% for most T2 values except for vial #2 (4.8%) and #9 (6.6%), indicating excellent intersession repeatability. Similarly, minimal intrasession variations were observed (CVs for T1 within 1% and T2 within 4.3%). Intra-session and inter-session temperature variations across the experiments were negligible (<0.5°C and <1°C, respectively), thus no temperature calibration was deemed necessary.

Figure 2C and 2D summarizes the mean T1 and T2 values in each session as compared to the manufacturer-provided reference data. While linear regression fitting revealed systematic biases, excellent intersession repeatability was observed as characterized by tightly clustered datapoints which overall deviated from the identity line. Strong linear associations between the DL-MUPA derivations and the manufacturer reference were observed for both T1 ($R^2$ = 0.998) and T2 ($R^2$ = 0.997).

**Brain Metastases Cohort**

Three patients with brain metastases were evaluated for NAWM and exhibited minimal variations with $|\Delta T1_{mean}|$<36 ms (4.9%), $|\Delta T2_{mean}|$<2 ms (6.1%) between different timepoints, except that variations >10% were observed in T1 for Patient B. wCV of NAWM was 1.7%-2.4% except for one outlier being 9.6% for T1 and 0.7%-3.4% for T2.

Eight lesions from three brain metastases Patients A-C underwent delta analysis. Table 2 summarizes detailed results of longitudinal changes and wCV of T1 and T2 within each ROI including the lesions and NAWM for patients A-C. Figure 3 highlights representative results of Patient A and B acquired over a 7-month timeframe. For Patient A in Fig 3A, NAWM histograms demonstrated reliable consistency for both T1 and T2 quantification ($|\Delta T1_{mean}|$<3.4%, $|\Delta T2_{mean}|$<1.3%). Within the left frontal lesion (GTV1), the 4-month PostTx T1 and T2 maps highlighted an increase in both T1 ($\Delta T1_{mean}$=13.7%) and T2 ($\Delta T2_{mean}$=17.8%) while 7-month PostTx QMRI maps and histograms remained consistent to 4-month PostTx. The other lesions GTV2-4 (not shown in the figure), located in left posterior parafalcine, left superior frontal gyrus and left occipital lobe, respectively, exhibited resolution (GTV2-3) or stable appearance (GTV4) at follow-ups, demonstrating near-stable or increasing T1 and heterogeneous T2 changes. Patient B (Fig 3B1-3) demonstrated higher variability of T1 quantitative metrics for NAWM compared to other patients in the cohort (PreTx with 3- and 6-month PostTx $|\Delta T1_{mean}|$ ~10-17%) while the latter two timepoints showed closer agreement to each other ($\Delta T1_{mean}$=8.6%). T2 quantification appeared less varied ($|\Delta T2_{mean}|$<6.1%). As shown in Figure 3B1, the left thalamus lesion (GTV1) exhibited T2 enhancement ($\Delta T2_{mean}$ =38.2%) 3-month PostTx, while central necrosis was further identified on the 6-month PostTx +C T1 weighted images that corresponded to notable enhancement on T1 and T2 maps ($\Delta T1_{mean}$=39.7%, $\Delta T2_{mean}$=41.4% compared to 3-month PostTx). Two other metastases of this patient were also identified as necrosis 6-month PostTx, with key results summarized in Figure 3B2-3. Comparing 6- with 3-month PostTx, the right precentral gyrus lesion (GTV2) demonstrated similar enhancement ($\Delta T1_{mean}$=17.6%, $\Delta T2_{mean}$=8.7%) while the left middle frontal gyrus lesion (GTV3) showed minimal changes ($\Delta T1_{mean}$=-4.6%, $\Delta T2_{mean}$=-7.1%).

**HN Cancer Cohort**

Three of the five HN cancer patients were analyzed for tumor and organs at risk response. Table 3 summarizes key results of longitudinal changes and wCV of T1 and T2 within each ROI including the tumor and OARs for Patients I, II and IV, and uninvolved masseter for all five patients.

Figure 4 summarizes key results of Patient I with Stage I (T1N1) SCC of the right base of tongue. The treatment plan was adapted MidTx due to substantial decrease in tumor size. This patient was imaged with the GE AIR Open RT suite and immobilized at each timepoint.

The uninvolved masseter remained stable (|$\Delta T1_{mean}$|<5.0%, |$\Delta T2_{mean}$|<5.6% across all timepoints). Longitudinal resolution of the tumor was reflected on qualitative images and DL-MUPA derived T1 and T2 maps. Within the GTV, T1 and T2 histograms showed slight changes while delta analysis showed an increase of ($\Delta T2_{mean}$=12.8%) PostTx. When considering the decreased residual disease volumes over time, the T1 and T2 histograms had a tendency skewed toward higher values while delta analysis revealed stable T1 but increased T2 PostTx ($\Delta T2_{mean}$=18.1%). Due to the extensive disease, the ipsilateral submandibular gland adjacent to the disease received a mean dose of 64 Gy with accompanied longitudinal increases in QMRI endpoints (MidTx $\Delta T1_{mean}$=12.3%, $\Delta T2_{mean}$=10.8%, further increasing to PostTx $\Delta T1_{mean}$=14.6%, $\Delta T2_{mean}$=34.5%). Similarly, the ipsilateral parotid showed increasing T1 PostTx ($\Delta T1_{mean}$=12.4%) but stable T2. By contrast, the distant contralateral submandibular gland with a mean dose of 33 Gy had near-stable metrics (|$\Delta T1_{mean}$|<7.3%, |$\Delta T2_{mean}$|<6.3%) and the contralateral parotid also demonstrated negligible variations (|$\Delta T1_{mean}$|<5.6%, |$\Delta T2_{mean}$|<4.6%). The patient's QOL survey at the end of treatment reported dry mouth and difficulty swallowing at the end of treatment.

Figure 5 highlights results from Patient II with Stage IV (T4N2) SCC of the right base of tongue who exhibited ~10 kg weight loss MidTx necessitating plan adaptation. This patient was imaged with the GE AIR Open RT suite at each timepoint but not immobilized due to claustrophobia. The uninvolved masseter remained stable except for slightly higher T2 variations at PostTx (|$\Delta T1_{mean}$|<5.7%, |$\Delta T2_{mean}$|<11.8%). While changes of the GTV at the base

of tongue were observed on the T2 map, changes in the mean T1 and T2 were generally negligible ($|\Delta T1_{mean}|<1.8\%$, $|\Delta T2_{mean}|<2.5\%$). With reduced RD volumes, RD histogram peaks shifted towards lower values characterized by median T1 and T2 changes (PostTx $\Delta T1_{median}=-5.3\%$, $\Delta T2_{median}=-11.3\%$) and longitudinal skewness changes were observed, while the mean T1 and T2 showed minimal changes ($|\Delta T1_{mean}|<2.1\%$, $|\Delta T2_{mean}|<4.4\%$). The submandibular glands demonstrated substantial longitudinal changes in quantitative MRI endpoints with MidTx to PostTX $\Delta T1_{mean}$=28.1% to 43.8%, $\Delta T2_{mean}$=33.9% to 61.3%, and $\Delta T1_{mean}$=17.9% to 28.9%, $\Delta T2_{mean}$=15.4% to 36.4% for the ipsilateral (receiving mean dose of 69 Gy) and contralateral submandibular glands, respectively. Longitudinal changes were also observed for the ipsilateral ($\Delta T1_{mean}$=10.2% to 26.6%, $\Delta T2_{mean}$=8.9% to 18.4%) and contralateral ($\Delta T1_{mean}$=7.0% to 22.3%, $\Delta T2_{mean}$=4.6% to 13.4%) parotids. In conjunction, the QOL survey at the end of treatment revealed the patient reported significant dry mouth and difficulty swallowing.

Figure 6 demonstrates results from Patient IV diagnosed with Stage I (T1N1) SCC of the left base of tongue (red) with significant nodal metastases (orange). The patient's plan was not adapted due to limited changes in anatomy during the treatment course. Because of replaced hardware during the course of the trial, the immobilized patient underwent data collection for the GE GEM RT Open suite PreTx whereas GE AIR Open RT suite was used MidTx and PostTx. For the uninvolved masseter, MidTx and PostTx T1 histograms showed excellent agreement but both deviated from PreTx T1 histogram ($|\Delta T1_{mean}|>10.5\%$ (118 ms)) while the T2 quantification was consistent ($|\Delta T2_{mean}|$<8.0% (2 ms)). The base of tongue tumor showed slight changes with regional enhancement observed on PostTx T1 and T2 maps (GTV1 $\Delta T1_{mean}$=7.6%, $\Delta T2_{mean}$=33.4%, RD1 $\Delta T1_{mean}$=9.9%, $\Delta T2_{mean}$=36.6%). The significant nodal metastases largely resolved PostTx, leading to substantial changes in QMRI endpoints (GTV2 $\Delta T1_{mean}$=-39.5%, $\Delta T2_{mean}$=-32.7%, RD2 $\Delta T1_{mean}$ = -29.2%, $\Delta T2_{mean}$=-35.3%). Both parotids remained stable. The ipsilateral submandibular gland adjacent to the disease (mean dose = 64 Gy) exhibited T1 enhancement PostTx ($\Delta T1_{mean}$=15.1%) while the distant contralateral one demonstrated less change ($|\Delta T1_{mean}|$<8.1%). For both submandibular glands, T2 histograms agreed between MidTx and PostTx with a visible shift from the Pre-Tx (Figure 6, bottom). QOL survey reponses for the patient revealed dry mouth and difficulty swallowing at the end of treatment was resolved 3-month PostTx.

All five HN cancer patients were analyzed for the uninvolved contralateral masseter which exhibited higher T1 variations when different coils were used within the same subject. For example, within a single subject with all 3 timepoints acquired with the same coil (GE AIR Open RT suite, 3 HN patients I-III), $|\Delta T1_{mean}|$<69 ms (7.0%) and wCV<3.7%. When two different coils were used across the same subject (2 HN patients IV, V), more variability was observed ($|\Delta T1_{mean}|$=118-142 ms (10.5%-12.6%) and wCV=6.9%-7.3%. T2 variations were more comparable between cases using the same or different coils within the same subject. When the same coil (GE AIR Open RT suite) was used across the same subject (n=3), $|\Delta T2_{mean}|$=0-4 ms (0.4%-17.8% due to extremely low absolute T2 value of masseter) and

wCV=3.4%-10.8%. When different coils were used (n=2), $|\Delta T2_{mean}|$=1-3 ms (4.5%-11.7%) and wCV=4.1%-8.2%.

## Discussion

Our work evaluated a novel mpMRI sequence, DL-MUPA, for its feasibility of radiation treatment response assessment. The phantom study revealed excellent repeatability of the sequence although a systematic bias was identified, suggesting reliability in longitudinal studies after the offset is addressed. When the same coil was used, stable longitudinal results were generally obtained in uninvolved normal tissue in brain and HN with the exception of a few outliers, suggesting promising within-subject repeatability consistent with the literatures.[12,35–38] Quantitative changes in GTVs corresponded to brain metastases resolution or necrosis and HN tumor resolution, respectively. T1 and T2 enhancement in parotids and submandibular glands adjacent to the disease coincided with salivary functional changes based on QOL surveys.

Despite systematic bias compared to the phantom reference values, our DL-MUPA phantom benchmarking demonstrated a strong association of both T1 and T2 derivation to the reference ($R^2$>0.997) and excellent longitudinal repeatability (intersession CV<0.9% for T1 and <6.6% for T2), suggesting reliability in longitudinal studies. Systematic bias of T1 and T2 quantification is often reported when benchmarking QMRI sequences using the phantom. In physiological ranges, Carr *et al.*[24] reported 2.0%-8.5% median bias of T1 and T2 quantification on a 3T MR in a 1-year phantom benchmarking of conventional sequences, and Nejad-Davarani *et al*.[12] reported 9.5% mean bias comparing the STAGE T1 quantification against inversion recovery for a 0.35T MR-linac. Nevertheless, excellent longitudinal repeatability and linearity were reported in the above studies (intersession CV <6% and $R^2$>0.997 reported in similar T1 and T2 range) [12,24], agreeing with our results. In the context of treatment response assessment, excellent repeatability is essential when evaluating longitudinal changes. However, systematic biases introduce challenges to accurately quantify QMRI change thresholds associated with treatment response, although characterizing temporal changes within a patient is one way to address this limitation. To generalize this to a multi-institutional setting, careful calibration will be required.[39]

In NAWM in the brain, excellent longitudinal agreement (<6.1%) in T1 and T2 values as well as low within-subject CV (<3.4%) were observed except for one outlier for T1 (Patient B, CV=9.6%). One SRS case (Patient B) exhibited T1 changes in the NAWM that may be attributed to patient- pathophysiological variations such as a pre-existing medical condition or undergoing concurrent treatments such as steroids, chemotherapy, and immunotherapy.[40] The MR-SIM underwent a software upgrade between timepoints for this which may have contributed some additional uncertainty as QMRI dependencies on software version have been reported.[41] Technically speaking, other contributors to T1 mapping variations include imperfections in the slice profile, B1 inhomogeneities, and inversion efficiency, which have been reported to potentially introduce daily variations to relaxometry quantification.[42] Nevertheless, all data obtained were comparable to what has been reported in the literature (CV=1.5%-12.8% for T1, ~5% for T2)[12,35,36]. For absolute NAWM quantification, our study showed T1 overestimation and T2 underestimation compared to literature,[43] consistent with the bias of higher T1 and lower T2 observed in the phantom benchmarking experiments.

For metastatic brain tumor response, limited data are available for comparison in the literature. Konar *et al*.[44] utilized synthetic MRI (MAGnetic resonance Image Compilation) derived T1 and T2 maps to characterize brain metastases on a 3T MR. Each patient was imaged only once, either before or after the treatment, thus no same-subject comparison was available. ROI was drawn excluding cystic and necrotic regions. They reported increases in both T1 ($\Delta$=18.4%) and T2 ($\Delta$=14.0%) comparing treated to untreated brain metastases. In our study, multiple resolved or stable lesions from one patient demonstrated increasing or near-stable T1 and heterogenous T2 changes, while one resolved lesion from another patient exhibited both T1 and T2 decreases. Another case, by contrast, showed two necrotic brain metastases that exhibited T1 and T2 increases ($\Delta T1_{mean}$=17.6%-39.7%, $\Delta T2_{mean}$=8.7%-41.4%) with substantial central enhancement comparing 6-month with 3-month PostTx. A third necrotic lesion for the same patient remained stable in the QMRI metrics. These preliminary results suggest that, with confirmation in a larger cohort, there may be opportunities to distinguish between stable, recurrent, and necrotic lesions.

In HN, uninvolved contralateral masseter showed reasonable T1 consistency when imaged with the same coil (CV<3.7%) while higher variations were found when imaged with different coils (CV>6.9%). T2 quantification was comparable between cases using the same or different coils with overall CV=3.4%-10.8%. Other studies have characterized muscle or other uninvolved normal tissue and found typical QMRI CVs on the order of 3-10% for T1rho ADC CV=5.7%-10.0% [37,38], comparable to our results. Similar to the NAWM, our study showed substantial T1 overestimation and T2 underestimation for uninvolved masseter compared to the literature,[45] consistent with bias in phantom.

In terms of HN tumor response, Bonate *et al*. acquired longitudinal T1 and T2 maps over HNSCC patients treated on a 1.5T MR-linac and reported significant longitudinal T2 increase.[10] Our study suggests similar findings. Within the base of tongue GTVs, T1 values showed marginal changes while T2 enhancement was more remarkable (PostTx $\Delta T2_{mean}$>12.8%) for two of the tumors. The resolved significant nodal metastasis exhibited considerable decreases in QMRI endpoints instead (PostTx $\Delta T1_{mean}$=-39.5%, $\Delta T2_{mean}$=-32.7%). Within RD, histograms reflected the changes of pathology volume and composition while delta analysis suggested T2 enhancement within the base of tongue lesions and QMRI decreases within the nodal disease.

In the salivary gland, Zhou *et al.* reported significant longitudinal increase of quantitative T2 in bilateral parotids receiving mean dose of 29 Gy for nasopharyngeal carcinoma patients.[46]

While literature investigating quantitative T1 or T2 changes of submandibular glands are limited, Doornaert *et al*. reported longitudinal ADC increases at ~10 fx MidTx and 2-3 months PostTx for bilateral submandibular glands receiving mean dose of 59-73 Gy.[47] In our study, patient-specific results were found where PostTx data with in-field parotids and submandibular glands demonstrated increases in T1 (>12.4% and >14.6%, respectively) and T2 (>13.4% and >34.5%, respectively) except for one near-field parotid showing stable T2. Patients with substantial T1 and T2 changes had correlative severe dry mouth based on QOL surveys, highlighting potential endpoints or predictors for functional sparing and adaptation.

DL-MUPA derives both T1 and T2 maps in a single scan (4 min 30 sec and 6 min 30 sec for brain and HN, respectively) with relatively high resolution needed for radiation therapy ($1.5^3$ and $1.5\times1.5\times2$ mm$^3$, respectively). As a comparison, a previously reported STAGE method took 10 minutes on a 0.35T MR-linac to generate PD, T1 and R2* maps ($232.5\times310\times192$ mm$^3$ FOV, $1\times1\times3$ acquisition resolution, 12.5% oversampling) while the standard IR method can take a few hours for T1 mapping of a single slice.[12] DL-MUPA also incorporates DL enhancement for noise and artifact reduction and sharpness improvement. Further acceleration can be achieved by optimizing further DL enhancement with trade-offs in image quality. One limitation of our work is DL-MUPA was not benchmarked against established standard T1 and T2 mapping methods such as VFA and IR. As a surrogate, we benchmarked against manufacturer-provided reference in the phantom which yielded reasonable results. Literature has highlighted benchmarking multi-parametric sequences against conventional

T1 (IR) and T2 (multi-echo spin-echo) sequences. While strong linear associations ($R^2$>0.99) were observed, biases were also reported when compared against T1(4-9%) and higher deviations were reported for T2 (12-15%) within physiological ranges.[48,49] Another limitation of this work is that only a single site was evaluated. Future expansion could be considered to include more sites for multi-institutional benchmarking using other comparison metrics such as the intraclass correlation coefficient.[50,51]

During our longitudinal clinical studies spanning 28 months, several uncertainties were involved that introduced challenges. First, the MR software was upgraded twice while the software versioning has shown to influence reconstruction algorithms and quantitative MRI.[52] However, our one-year phantom benchmarking using the brain coil demonstrated excellent repeatability of DL-MUPA quantitative results despite different software versions. In addition, a new HN coil was implemented during the course of study leading to use of different coils at different timepoints in the analysis, including within a single subject. Multi-center studies have demonstrated that different radiofrequency coils have an impact in QMRI even when the same phantom has been used.[50,53] This uncertainty likely contributes to our finding that higher T1 variations in the uninvolved normal masseter were observed when different HN coils were used across timepoints within the same subject. For a coil comparison experiment performed using the same phantom and MR-SIM, DL-MUPA demonstrated close agreement (difference <5%) and strong linear association ($R^2$>0.99) in the physiological ranges between using the two HN coils. However, higher discrepancies (~11%) were observed in some T2 values (Supplementary material, Figure S1). Other sources of uncertainty include potential B1 field inhomogeneities, which contribute to daily variations [42] and are more prominent *in vivo* than in phantom.[54] Another challenge introduced was that some HN imaging timepoints did not use immobilization devices due to claustrophobia encountered during the scan. Similarly, brain patients were not immobilized as there is a physician preference to use a dedicated head coil for brain imaging that is not compatible with immobilization devices. These practical considerations are inherent in multi-year clinical trials that capture real-world data. Validation in a larger cohort would potentially help separate or mitigate such uncertainties.

Our work showed that DL-MUPA demonstrated excellent in vitro repeatability while yielding promising *in vivo* repeatability and tumor change detectability, suggesting promise in deriving QMRI biomarkers correlated to treatment response. With validation in a larger cohort, associations with clinical outcome data such as survival, patient reported outcomes, and normal tissue toxicities can be more firmly established to enable treatment response prediction and treatment adaptation guidance.[55] Other potential applications of DL-MUPA include generating quantitative PD maps from the PD weighted images by local normalization after correcting B1 transmit and receive contributions, and generating synthetic CT from the Zero TE PD weighted images to support MR-only radiation treatment workflow[18], which were beyond the scope of the current study.

## Conclusion

Our results demonstrate the feasibility of applying DL-MUPA for assessing longitudinal treatment response. DL-MUPA was stable in phantom and in uninvolved regions *in vivo*, suggesting excellent repeatability, while lesion and organs abutting targets showed demonstrable changes highlighting the sensitivity of our technique. With further confirmation and coupled with clinical outcome information, further correlation with T1 and T2 via DL-MUPA can be established to further identify actionable endpoints for functional treatment adaptation.

## Figure

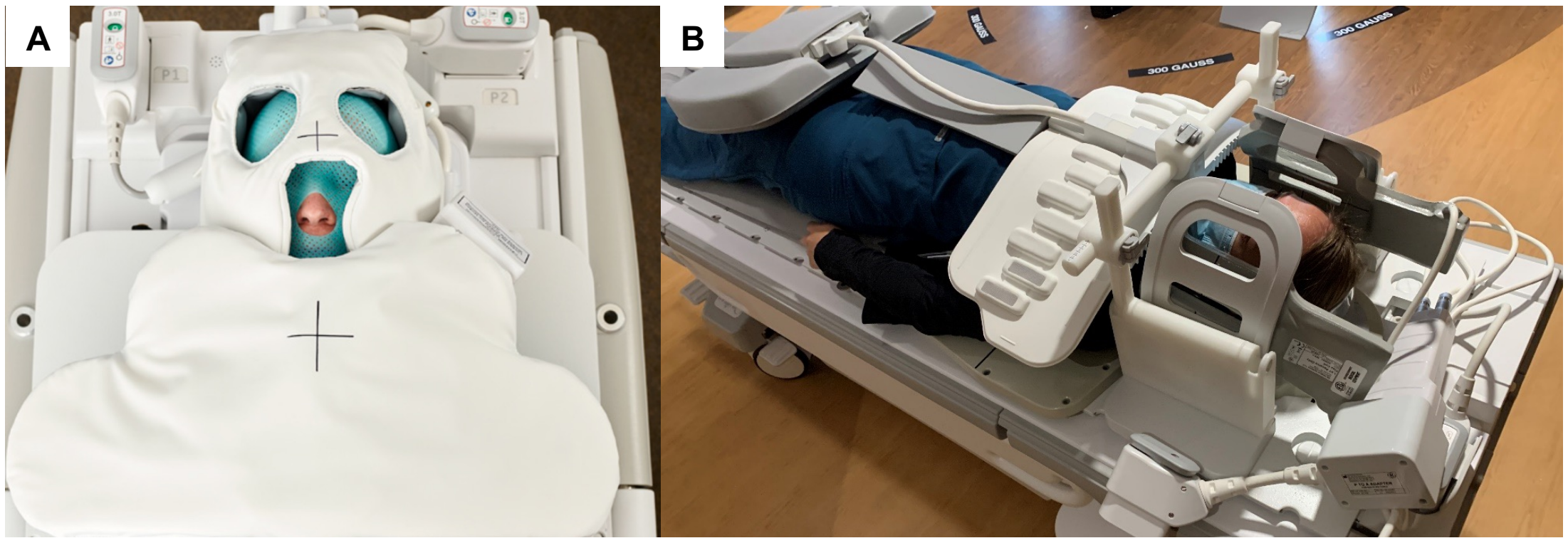

Fig. 1. Head and neck coils used in patient study including (A) 32-channel GE AIR Open RT suite and (B) 30-channel GE GEM RT Open suite.

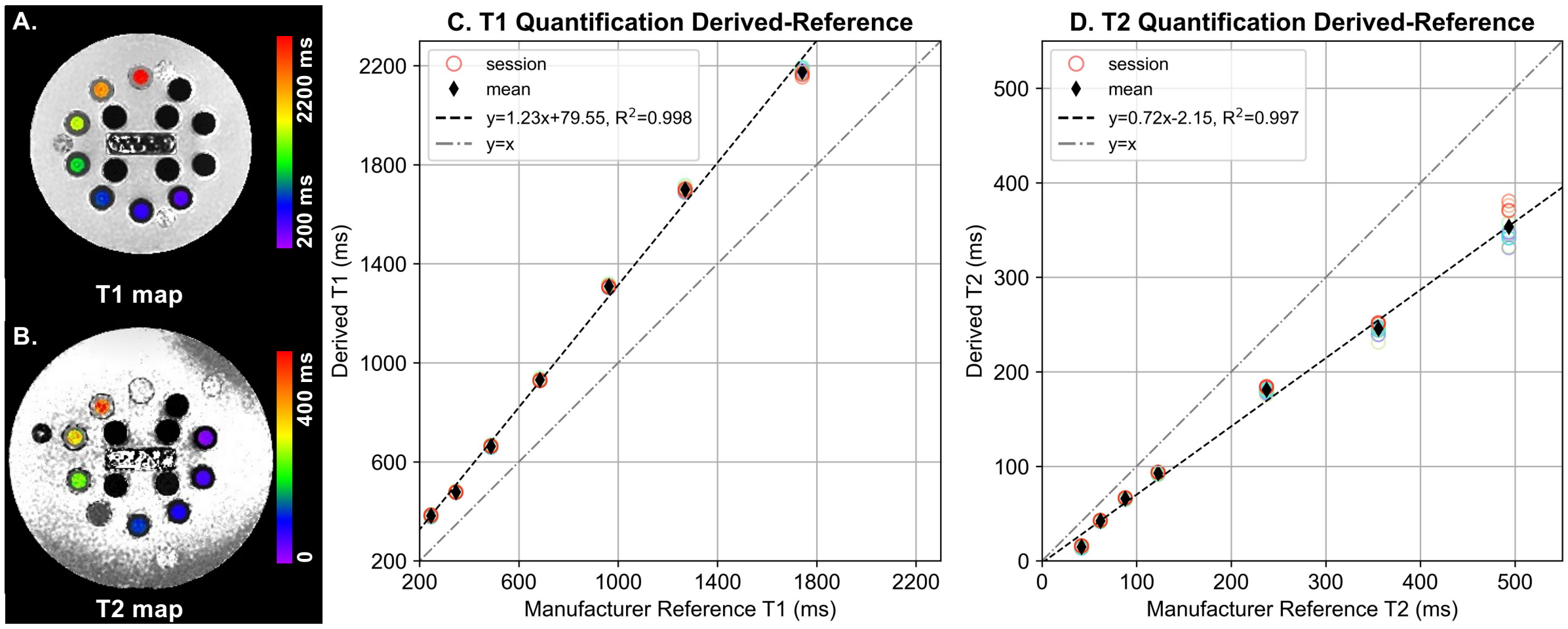

Fig. 2. Example (A) T1 and (B) T2 map of corresponding vial array derived on the phantom by DL-MUPA. Derived (C) T1 and (D) T2 values are plotted against manufacturer-provided reference with each circle standing for mean derived value of five acquisitions in a single session and each thin diamond standing for mean derived value over all acquisitions (12×5). Least square linear regression lines were plotted in black dash line and identity lines were plotted in grey dash-dot line.

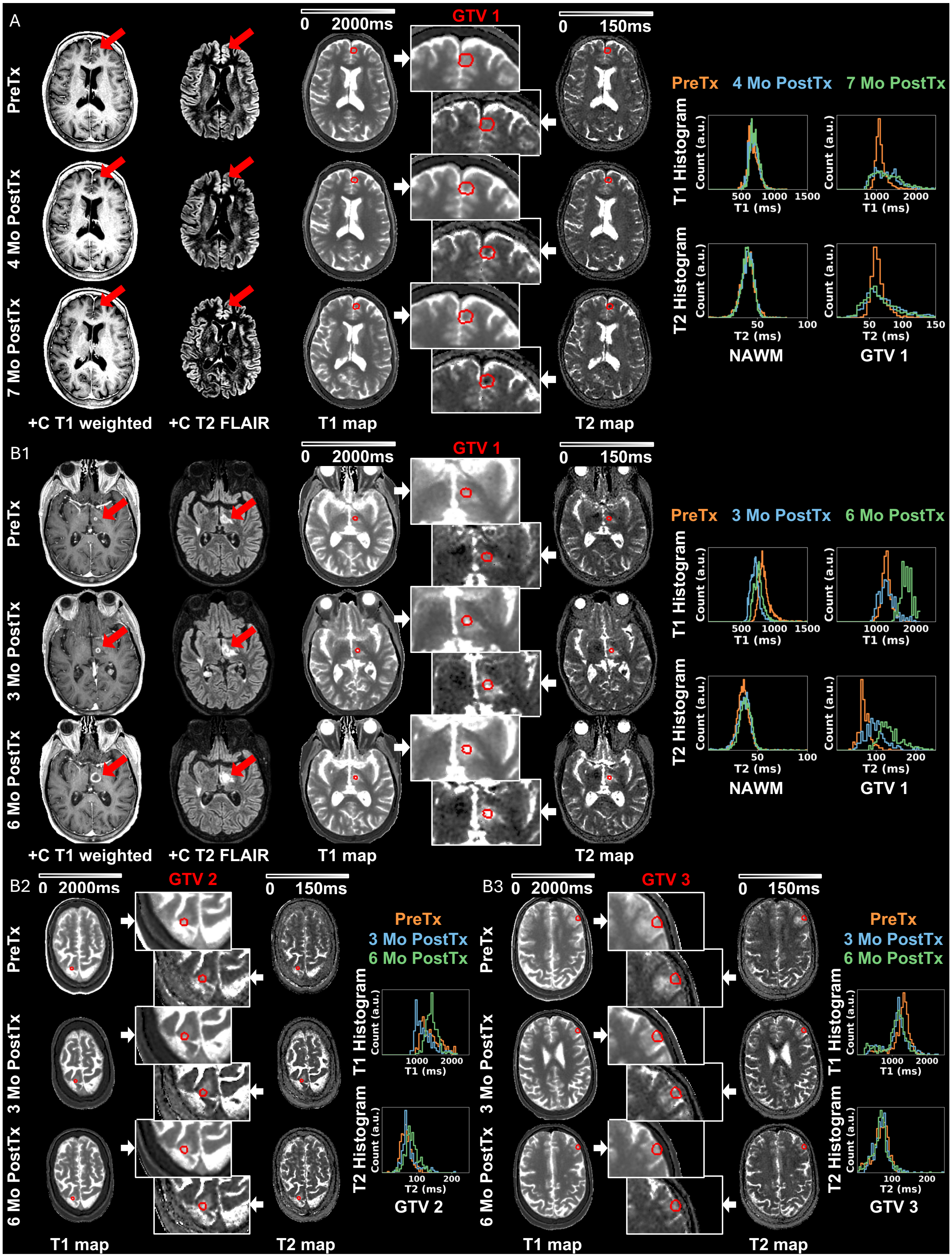

A
PreTx
4 Mo PostTx
7 Mo PostTx
+C T1 weighted
+C T2 FLAIR
0 2000ms
GTV 1
0 150ms
T1 map
T2 map
PreTx 4 Mo PostTx 7 Mo PostTx
T1 Histogram
T2 Histogram
Count (a.u.)
T1 (ms)
T2 (ms)
NAWM
GTV 1
B1
PreTx
3 Mo PostTx
6 Mo PostTx
+C T1 weighted
+C T2 FLAIR
T1 map
T2 map
PreTx 3 Mo PostTx 6 Mo PostTx
NAWM
GTV 1
B2
GTV 2
B3
GTV 3

Fig. 3. Longitudinal post-contrast T1 weighted images and T2 FLAIR, DL-MUPA derived T1 and T2 maps (global and magnified to metastases), and quantitative T1 and T2 histograms of normal appearing white matter (NAWM) and gross tumor volume (GTV), (A) presenting one resolved lesion for brain metastases Patient A and (B1-3) presenting three necrotic lesions for Patient B.

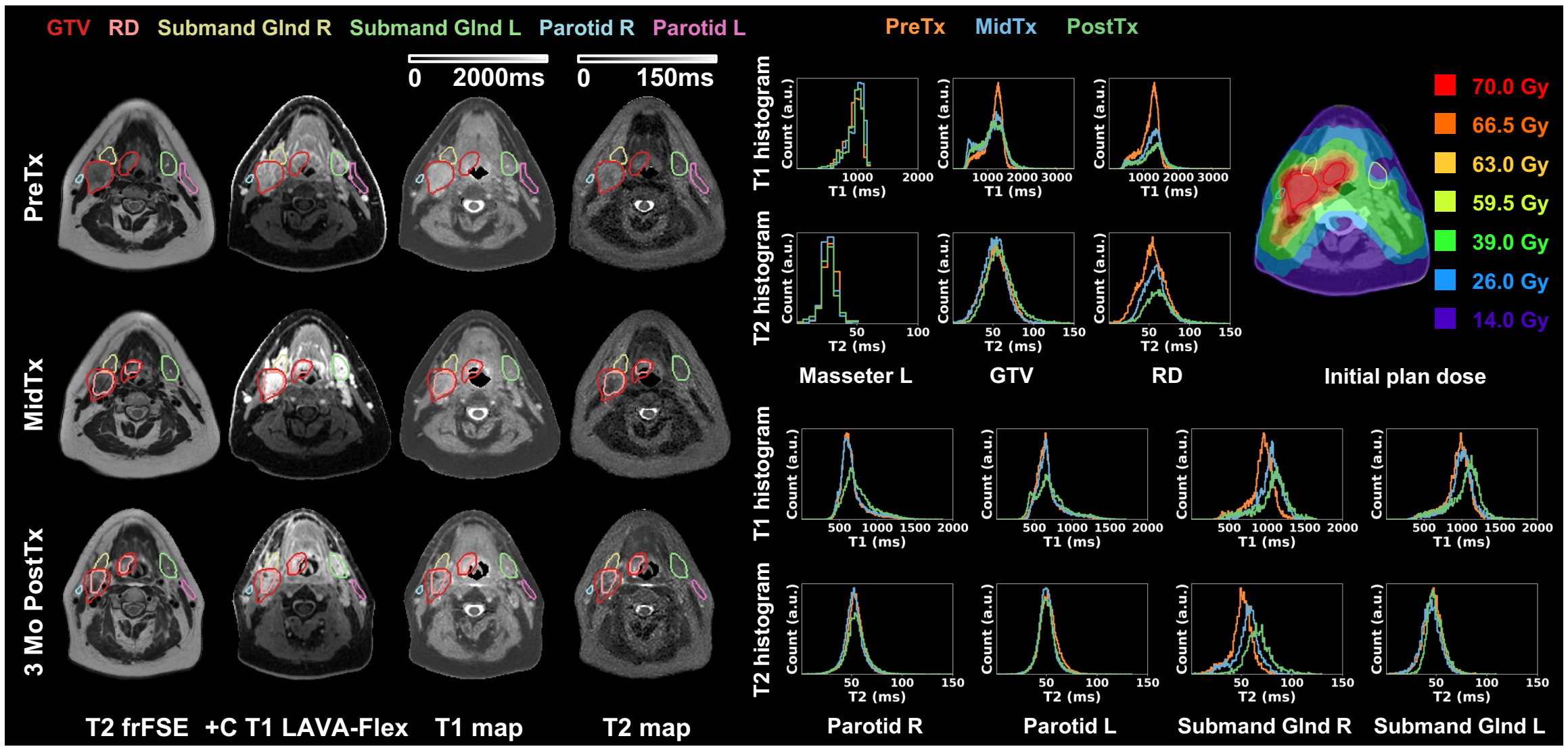

Fig. 4. Longitudinal T2 frFSE, post-contrast T1 LAVA-Flex, DL-MUPA derived T1 and T2 maps, quantitative T1 and T2 histograms including uninvolved masseter, gross tumor volume (GTV), residual disease (RD), parotids and submandibular glands, and initial plan dose for Head and Neck cancer patient I.

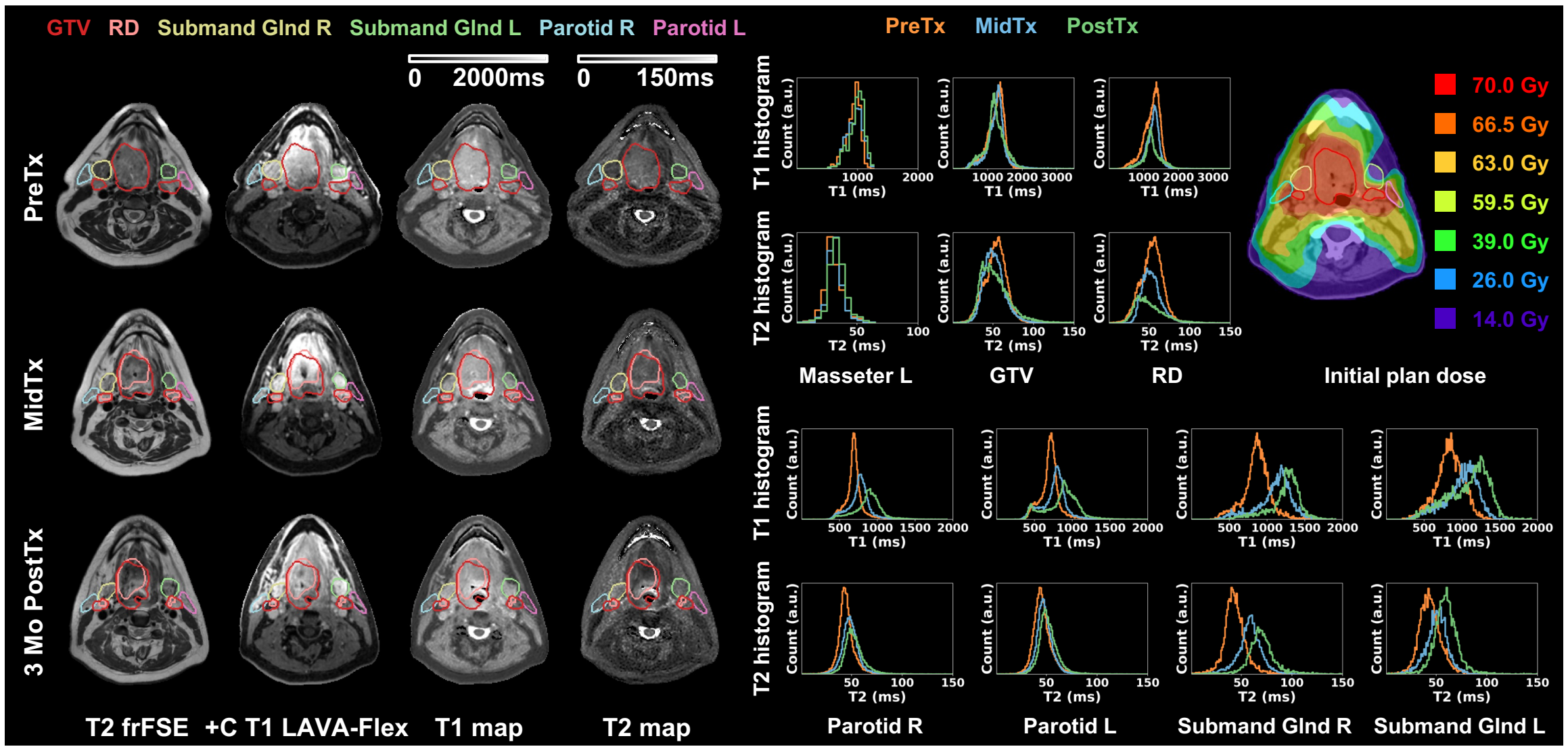

Fig. 5. Longitudinal T2 frFSE, post-contrast T1 LAVA-Flex, DL-MUPA derived T1 and T2 maps, quantitative T1 and T2 histograms including uninvolved masseter, gross tumor volume (GTV), residual disease (RD), parotids and submandibular glands, and initial plan dose for Head and Neck cancer patient II, exhibiting considerable changes in parotids and submandibular glands.

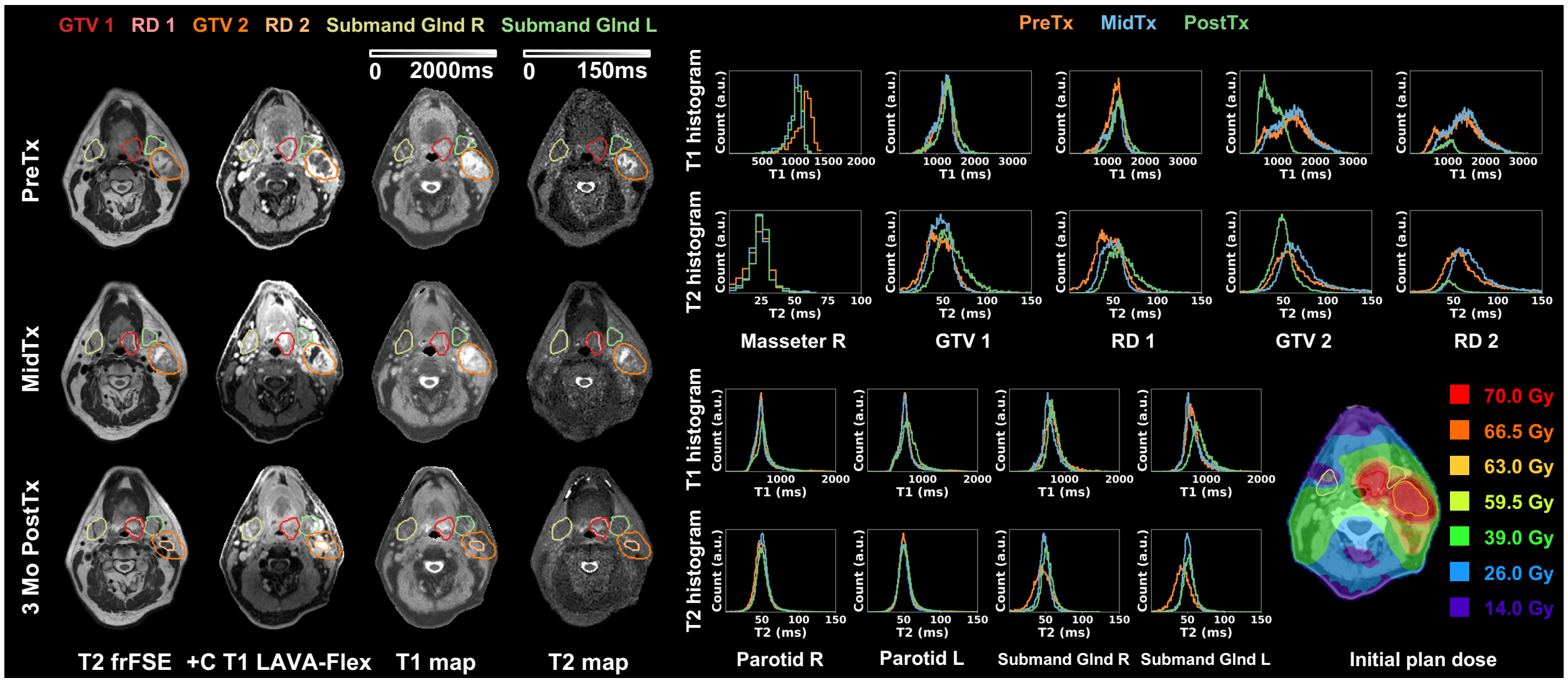

Fig. 6. Longitudinal T2 frFSE, post-contrast T1 LAVA-Flex, DL-MUPA derived T1 and T2 maps, quantitative T1 and T2 histograms including uninvolved masseter, gross tumor volume (GTV), residual disease (RD), parotids and submandibular glands, and initial plan dose for Head and Neck cancer patient IV demonstrating base of tongue primary tumor (GTV/RD1) and resolution of significant nodal metastases (GTV/RD2)

Table 1. Summarized results of 12-session phantom benchmarking, each session including 5 consecutive scans. T1 and T2 values within physiologically relevant range were reported for the mean, intersession standard deviation (SD), and intersession coefficient of variance (CV) evaluating the repeatability across the 12 sessions. The intra-session CV evaluated the repeatability of the 5 consecutive acquisitions within a single session. The bias in absolute value and percentage are summarized comparing to manufacturer-provided reference values as shown.

| | Vial | Mean (ms) | Intersession SD (ms) | Intersession CV (%) | Intrasession CV (Range, %) | Phantom Reference (ms) | Bias (ms) | Bias (%) |
|---|---|---|---|---|---|---|---|---|
| T1 Quantification (N=12) | 1 | 2175.6 | 12.1 | 0.6% | [0.1%, 0.9%] | 1741.3 | 434.3 | 24.9% |
| | 2 | 1700.4 | 7.6 | 0.4% | [0.1%, 0.6%] | 1269.8 | 430.6 | 33.9% |
| | 3 | 1309.4 | 4.5 | 0.3% | [0.1%, 0.5%] | 962.0 | 347.4 | 36.1% |
| | 4 | 931.1 | 3.4 | 0.4% | [0.0%, 0.5%] | 684.7 | 246.3 | 36.0% |
| | 5 | 662.8 | 3.6 | 0.5% | [0.1%, 0.8%] | 487.4 | 175.5 | 36.0% |
| | 6 | 478.0 | 2.1 | 0.4% | [0.2%, 1.0%] | 346.4 | 131.6 | 38.0% |
| | 7 | 382.3 | 3.3 | 0.9% | [0.1%, 1.0%] | 245.8 | 136.6 | 55.6% |
| T2 Quantification (N=12) | Vial | Mean (ms) | Intersession SD (ms) | Intersession CV (%) | Intrasession CV (Range, %) | Phantom Reference (ms) | Bias (ms) | Bias (%) |
| | 2 | 353.6 | 17.0 | 4.8% | [0.4%, 2.7%] | 493.4 | -139.8 | -28.3% |
| | 3 | 245.8 | 5.9 | 2.4% | [0.6%, 1.8%] | 355.5 | -109.7 | -30.9% |
| | 4 | 181.2 | 2.8 | 1.5% | [0.2%, 1.1%] | 237.3 | -56.1 | -23.6% |
| | 6 | 92.8 | 0.9 | 1.0% | [0.3%, 0.6%] | 122.7 | -29.9 | -24.4% |
| | 7 | 65.9 | 0.7 | 1.0% | [0.3%, 0.6%] | 88.0 | -22.1 | -25.1% |
| | 8 | 42.3 | 0.5 | 1.3% | [0.2%, 0.7%] | 61.6 | -19.4 | -31.4% |
| | 9 | 14.8 | 1.0 | 6.6% | [1.3%, 4.3%] | 41.8 | -27.0 | -64.6% |

Table 2. Longitudinal changes of mean T1 and T2 values and within-subject coefficient of variance (wCV) for each analyzed region of interest (ROI) for the three brain metastases patients. PostTreatment (PostTx) scans were acquired every 3-4 months after stereotactic radiosurgery.

| **Uninvolved normal tissue** | | | | | | | | | | |
|---|---|---|---|---|---|---|---|---|---|---|
| **Patient** | **ROI** | **PreTx $T1_{mean}$ (ms)** | **1st PostTx $\Delta T1_{mean}$ (ms (%))** | **2nd PostTx $\Delta T1_{mean}$ (ms (%))** | **T1 wCV (%)** | **PreTx $T2_{mean}$ (ms)** | **1st PostTx $\Delta T2_{mean}$ (ms (%))** | **2nd PostTx $\Delta T2_{mean}$ (ms (%))** | **T2 wCV (%)** |
| **A** | **NAWM** | 678 | 9 (1.3%) | 23 (3.4%) | 1.7 | 39 | 1 (1.3%) | 0 (0.1%) | 0.7 |
| **B** | **NAWM** | 855 | -148 (-17.3%) | -87 (-10.2%) | 9.6 | 36 | 0 (0.2%) | 2 (6.1%) | 3.4 |
| **C** | **NAWM** | 740 | 19 (2.6%) | 36 (4.9%) | 2.4 | 37 | 0 (-1.2%) | 1 (1.7%) | 1.5 |
| **Tumor** | | | | | | | | | |
| **Patient** | **ROI** | **PreTx $T1_{mean}$ (ms)** | **1st PostTx $\Delta T1_{mean}$ (ms (%))** | **2nd PostTx $\Delta T1_{mean}$ (ms (%))** | **T1 wCV (%)** | **PreTx $T2_{mean}$ (ms)** | **1st PostTx $\Delta T2_{mean}$ (ms (%))** | **2nd PostTx $\Delta T2_{mean}$ (ms (%))** | **T2 wCV (%)** |
| **A** | **GTV 1** | 1129 | 155 (13.7%) | 193 (17.1%) | 8.2 | 65 | 12 (17.8%) | 11 (16.2%) | 8.8 |
| | **GTV 2** | 955 | 44 (4.6%) | 45 (4.7%) | 2.6 | 51 | 5 (10.3%) | 3 (5.2%) | 4.9 |
| | **GTV 3** | 1433 | 99 (6.9%) | 160 (11.2%) | 5.3 | 153 | -5 (-3.0%) | 22 (14.2%) | 8.9 |
| | **GTV 4** | 926 | 70 (7.6%) | 74 (8.0%) | 4.3 | 43 | -5 (-12.7%) | -6 (-13.7%) | 8.3 |
| **B** | **GTV 1** | 1257 | 8 (0.6%) | 510 (40.6%) | 20.4 | 72 | 27 (38.2%) | 68 (95.4%) | 33.2 |
| | **GTV 2** | 1399 | -181 (-13.0%) | 33 (2.3%) | 8.5 | 69 | 10 (13.8%) | 16 (23.7%) | 10.6 |
| | **GTV 3** | 1249 | -124 (-9.9%) | -176 (-14.1%) | 7.9 | 67 | 4 (5.8%) | -1 (-1.7%) | 3.9 |
| **C** | **GTV 1** | 1007 | -33 (-3.2%) | -159 (-15.8%) | 8.9 | 53 | -5 (-10.2%) | -13 (-25.1%) | 14.3 |

Table 3. Longitudinal changes of mean T1 and T2 values and within-subject coefficient of variance (wCV) for each analyzed region of interest (ROI) for the five head and neck (HN) cancer patients. Masseter was defined as the uninvolved control ROI to assess sequence stability. Patients I and II had squamous cell cancer (SCC) of the right base of tongue, while Patient IV had SCC of the left based of tongue (GTV 1) and left nodal metastases (GTV 2). Residual disease (RD) volumes were defined on the MidTx and PostTx datasets to reflect the underlying anatomy at the given timepoint.

**Uninvolved normal tissue**

| Patient | ROI | PreTx $T1_{mean}$ (ms) | MidTx $\Delta T1_{mean}$ (ms (%)) | PostTx $\Delta T1_{mean}$ (ms (%)) | T1 wCV (%) | PreTx $T2_{mean}$ (ms) | MidTx $\Delta T2_{mean}$ (ms (%)) | PostTx $\Delta T2_{mean}$ (ms (%)) | T2 wCV (%) |
|---|---|---|---|---|---|---|---|---|---|
| I | Masseter | 934 | 47 (5.0%) | 18 (1.9%) | 2.5 | 28 | -2 (-5.6%) | -2 (-5.6%) | 3.4 |
| II | Masseter | 931 | 22 (2.4%) | 53 (5.7%) | 2.8 | 31 | 1 (4.5%) | 4 (11.8%) | 5.6 |
| III | Masseter | 994 | -21 (-2.1%) | -69 (-7.0%) | 3.7 | 22 | 0 (-0.4%) | -4 (-17.8%) | 10.8 |
| IV | Masseter | 1124* | -142 (-12.6%) | -118 (-10.5%) | 7.3 | 22* | 1 (6.7%) | 2 (8.0%) | 4.1 |
| V | Masseter | 1115* | 10 (0.9%)* | -124 (-11.1%) | 6.9 | 23* | 3 (11.7%)* | -1 (-4.5%) | 8.2 |

**Tumor and nearby organs at risk**

| Patient | ROI | PreTx $T1_{mean}$ (ms) | MidTx $\Delta T1_{mean}$ (ms (%)) | PostTx $\Delta T1_{mean}$ (ms (%)) | T1 wCV (%) | PreTx $T2_{mean}$ (ms) | MidTx $\Delta T2_{mean}$ (ms (%)) | PostTx $\Delta T2_{mean}$ (ms (%)) | T2 wCV (%) |
|---|---|---|---|---|---|---|---|---|---|
| I | GTV | 1166 | -76 (-6.5%) | -46 (-3.9%) | 3.4 | 53 | -1 (-1.9%) | 7 (12.8%) | 7.7 |
| | RD | 1166 | 56 (4.8%) | 66 (5.6%) | 2.9 | 53 | 4 (6.9%) | 10 (18.1%) | 8.4 |
| | Parotid R | 660 | 12 (1.9%) | 82 (12.4%) | 6.4 | 53 | -1 (-2.5%) | 1 (2.4%) | 2.5 |
| | Parotid L | 674 | 22 (3.3%) | 38 (5.6%) | 2.8 | 52 | -2 (-4.6%) | -2 (-3.1%) | 2.4 |
| | Submand Glnd R | 926 | 113 (12.3%) | 135 (14.6%) | 7.2 | 50 | 5 (10.8%) | 17 (34.5%) | 15.3 |
| | Submand Glnd L | 955 | 30 (3.2%) | 69 (7.3%) | 3.5 | 46 | -2 (-3.9%) | 3 (6.3%) | 5.1 |
| II | GTV | 1237 | 13 (1.1%) | -23 (-1.8%) | 1.5 | 53 | -1 (-2.5%) | 1 (2.1%) | 2.3 |
| | RD | 1237 | 26 (2.1%) | -3 (-0.3%) | 1.3 | 53 | 0 (0.3%) | -2 (-4.4%) | 2.7 |
| | Parotid R | 694 | 71 (10.2%) | 185 (26.6%) | 12 | 44 | 4 (8.9%) | 8 (18.4%) | 8.4 |
| | Parotid L | 705 | 50 (7.0%) | 157 (22.3%) | 10.4 | 46 | 2 (4.6%) | 6 (13.4%) | 6.4 |
| | Submand Glnd R | 859 | 241 (28.1%) | 376 (43.8%) | 17.9 | 43 | 15 (33.9%) | 26 (61.3%) | 23.3 |
| | Submand Glnd L | 838 | 150 (17.9%) | 242 (28.9%) | 12.6 | 43 | 7 (15.4%) | 16 (36.4%) | 15.6 |
| IV | GTV 1 | 1187* | -21 (-1.8%) | 90 (7.6%) | 4.9 | 44* | 2 (4.6%) | 15 (33.4%) | 16.1 |
| | RD 1 | 1187* | 15 (1.3%) | 117 (9.9%) | 5.2 | 44* | 4 (8.3%) | 16 (36.6%) | 16.7 |
| | GTV 2 | 1371* | 38 (2.8%) | -541 (-39.5%) | 26.9 | 72* | 0 (0.5%) | -24 (-32.7%) | 21.3 |
| | RD 2 | 1371* | 82 (6.0%) | -400 (-29.2%) | 20.4 | 72* | 3 (3.6%) | -25 (-35.3%) | 24 |
| | Parotid R | 692* | -30 (-4.3%) | 23 (3.3%) | 3.8 | 49* | 2 (3.3%) | 2 (3.3%) | 1.9 |
| | Parotid L | 728* | -29 (-4.0%) | 24 (3.3%) | 3.6 | 52* | -2 (-4.0%) | 0 (0.1%) | 2.4 |
| | Submand Glnd R | 797* | -64 (-8.1%) | 10 (1.2%) | 5.2 | 45* | 2 (5.4%) | 7 (15.3%) | 7.3 |
| | Submand Glnd L | 772* | -7 (-0.9%) | 117 (15.1%) | 8.6 | 42* | 8 (18.7%) | 10 (23.2%) | 10.8 |

* Acquired using GE GEM RT Open suite; all other data were acquired with GE AIR Open RT suite.

## Supplement Document

To determine the impact of the different HN coils on the quantitative DL-MUPA derived T1 and T2 results, a coil comparison was performed using the same phantom and 1.5T GE MR scanner as described in the Methods. The NIST-ISMRM phantom was scanned with DL-MUPA using the two HN coils (GE GEM RT Open suite and GE AIR Open RT suite). DL-MUPA acquisition was repeated five times consecutively for each coil and the phantom temperature was monitored throughout the data acquisition. T1 and T2 quantifications were analyzed following the same methodology described in the Method section, under Phantom Benchmarking, Image Processing and Quantitative Analysis.

Due to a recent scanner platform upgrade to Version MR30.1 R04, a constancy check was performed by scanning the phantom with the brain coil following the above methodology and data were compared to previous phantom benchmarking. Minimal temperature variations were found within 0.6 °C thus no temperature calibration was deemed necessary. Within physiological ranges (reference T1 [246-1741 ms]), minimal differences were found in T1 quantification (<1.3%) using the brain coil as compared to the prior longitudinal phantom benchmarking. However, for reference T2 values in physiological ranges of 42-493 ms, slightly larger variations in T2 values (up to 7.4%) were observed from the prior measurements and similar magnitude to the intersession CV of up to 6.6% for T2.

Comparing phantom measurements of the two HN coils (e.g., GE GEM RT Open suite and GE AIR Open RT suite) within the same physiological range as depicted in Figure S1, close agreement (<5% difference) and strong linear association ($R^2$>0.99) were observed for both T1 and T2. However, larger discrepancies were observed for the T2 vials of ~350 ms and ~30 ms where differences were on the order of ~11%. For a general comparison, our prior phantom benchmarking also demonstrated higher variations for similar T2 values (CV=4.8%-6.6%) as compared with the other T2 vials that exhibited a CV<2.4%. Within the phantom QMRI values similar to that expected for the masseter examined in the HN cohort (T1~900-1100 ms, T2~25 ms), the HN coil comparison showed excellent T1 agreement (<1%) with higher T2 differences (~11%).

While the phantom experiment suggested consistency in T1 values between the two HN coils, further in vivo investigation is warranted. Other studies in the literature have reported mixed results of in vivo QMRI agreement using different coils.[1,2] Other confounding factors such as correcting for B1 inhomogeneities can be explored in future work to further reduce uncertainties of daily variations in QMRI.[3]

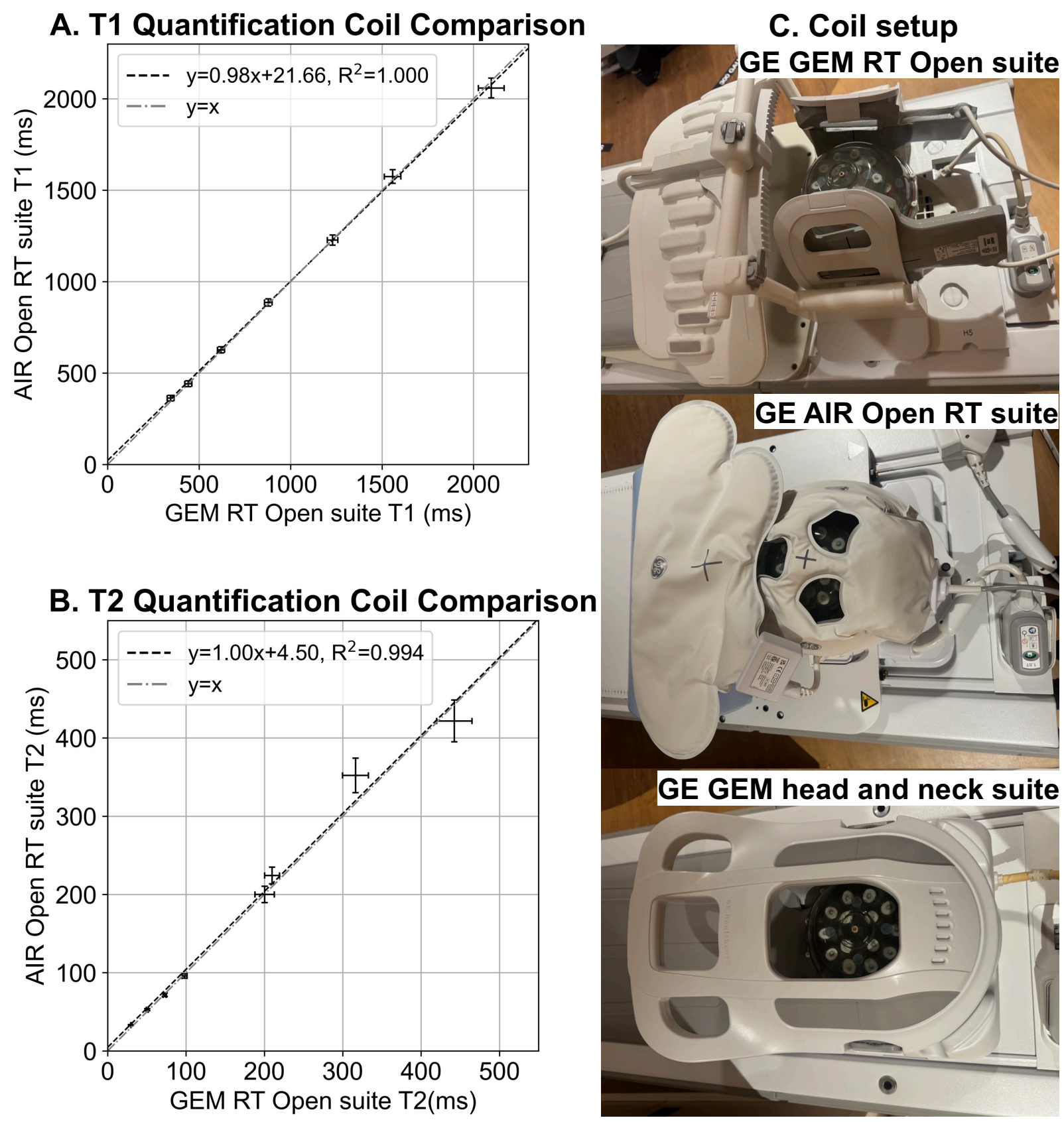

Figure S1. Comparison of DL-MUPA (A) T1 and (B) T2 quantifications using the GE GEM RT Open suite coil (x-axis) and the GE AIR Open RT suite coil (y-axis). The center and width of each cross represent mean and intra-session standard deviation of quantification of each vial, respectively. (C) Different coil configurations used in the phantom study.